\documentclass[runningheads]{llncs}

\usepackage[T1]{fontenc}
\usepackage{tabularx}
\usepackage{wrapfig}
\usepackage{graphicx}%
\usepackage{multirow}%
\usepackage{amsmath,amssymb,amsfonts}%
\usepackage{amsthm}%
\usepackage{mathrsfs}%
\usepackage[title]{appendix}%
\usepackage{xcolor}%
\usepackage{textcomp}%
\usepackage{manyfoot}%
\usepackage{booktabs}%
\usepackage{algorithm}%
\usepackage{algorithmicx}%
\usepackage{algpseudocode}%
\usepackage{listings}%

\begin{document}
\title{Enhanced Interpretable Knowledge Tracing \textit{for}  \\ Students' Performance Prediction with Human-understandable Feature Space}
\titlerunning{Enhanced Interpretable Knowledge Tracing}
%
\author{Sein Minn\inst{1,2} \and Roger Nkambou\inst{2} }
%
%

\institute{Asian Institute of Technology, Thailand\\ \email{\{sein-minn\}@ait.asia}\\ AI Research Center - Université du Québec à Montréal, Montreal, Canada\\ \email{\{nkambou.roger\}@uqam.ca}}
\maketitle              
\begin{abstract}
Knowledge Tracing (KT) plays a central role in assessing students' skill mastery and predicting their future performance. While deep learning-based KT models achieve superior predictive accuracy compared to traditional methods, their complexity and opacity hinder their ability to provide psychologically meaningful explanations. This disconnect between model parameters and cognitive theory poses challenges for understanding and enhancing the learning process, limiting their trustworthiness in educational applications. To address these challenges, we enhance interpretable KT models by exploring human-understandable features derived from students’ interaction data. By incorporating additional features, particularly those reflecting students' learning abilities, our enhanced approach improves predictive accuracy while maintaining alignment with cognitive theory. Our contributions aim to balance predictive power with interpretability, advancing the utility of adaptive learning systems.

\end{abstract}
\section{Introduction}
Adaptive learning systems have revolutionized modern education by dynamically adjusting to each learner’s requirements in real time\cite{fu2013recommender,minn2022ai,minn2025understanding}. However, implementing such personalized learning at scale presents significant challenges, where intelligent tutoring systems must effectively assess and adapt to the learning trajectories of tens or even hundreds of thousands of students~\cite{minn2022interpretable}.

KT models are the backbone of intelligent tutoring systems, driving personalized content delivery and feedback. Artificial Intelligence (AI) plays a critical role in this process, with KT models using machine learning techniques to infer a student's conceptual and procedural knowledge based on their task performance~\cite{corbett1994knowledge,minn2016refinement}. Among KT methods, Deep Knowledge Tracing (DKT)~\cite{piech2015deep} stands out for its use of Neural Networks (NNs), which model student learning over time and deliver superior predictive performance, highlighting the transformative impact of deep learning in education~\cite{piech2015deep,minn2019dynamic,minn2020bkt}.

Bayesian Knowledge Tracing (BKT)~\cite{corbett1994knowledge}, an earlier and interpretable KT model, utilizes a Hidden Markov Model to track a student’s knowledge state. In contrast, DKT overcomes this limitation by using NNs to model cross-skill dynamics, improving predictive accuracy. Despite its improved performance, DKT's reliance on complex hidden layers and large numbers of parameters reduces its interpretability, making it less transparent and trustworthy for educators. The lack of interpretability in models like DKT highlights the need for methods that combine accurate predictions with clear, actionable insights. Toward a transparent, psychologically meaningful model, we aim to extend IKT~\cite{minn2022interpretable}, focusing on simplicity while maintaining comparable predictive performance.

\section{Exploring human-understandable feature space}

Drawing inspiration from the Interpretable Knowledge Tracing (IKT) model~\cite{minn2022interpretable}, we propose an Enhanced Interpretable Knowledge Tracing (EIKT) model that introduces a richer and more interpretable feature space through advanced feature extraction. The IKT model predicts a student’s future responses using three core latent features: individual skill ID, skill mastery, and problem difficulty. Beyond this foundation, EIKT incorporates enhanced ability profiling to explore the interplay between item difficulty and skill correctness.

\subsubsection{Skill Mastery}
In a typical BKT-based learning environment~\cite{corbett1994knowledge}, a student's skill mastery is dynamically updated with each item's response.  BKT operates using four key probabilities: \textbf{Initial Knowledge ($P(L_0)$):} The probability that a student already knows the skill before any practice; \textbf{Learning ($P(T)$):} The probability that the student will transition from not knowing the skill to knowing it after an opportunity to practice; \textbf{Guess ($P(G)$):} The probability that the student correctly answers an item despite not having mastered the skill; \textbf{Slip ($P(S)$):} The probability that the student incorrectly answers an item despite having mastered the skill.

The probability that a student knows the skill at any given time $t$ is continuously updated based on their observed outcomes, recorded as either successful ($\text{Obs}=1$) or unsuccessful ($\text{Obs}=0$). This iterative update mechanism allows BKT to provide a dynamic and data-driven assessment of a student's evolving knowledge state.

\begin{equation} \label{equ:corr}
P(L_{t}|1) = \frac{P(L_{t})(1-P(S))}{P(L_{t})(1-P(S))+(1-P(L_{t}))P(G)}
\end{equation}

\begin{equation} \label{equ:incorr}
P(L_{t}|0) = \frac{P(L_{t}) P(S)}{P(L_{t})P(S)+(1-P(L_{t}))(1-P(G))}
\end{equation}

\begin{equation} \label{equ:action}
P(L_{t+1}) = P(L_{t}|obs)+ (1-P(L_{t}|obs))P(T)
\end{equation}

 \begin{equation} 
\textbf{skill mastery}(s_t)=\delta(P(L_{t}),s_t)
\end{equation}

In this context, $\delta(P(L_{t}), s_t)$ represents a function that maps the mastery level of a specific skill $s_t$ at the current timestamp, based on the student's cumulative interaction history.

\subsubsection{Problem Difficulty}\label{sec:difficulty}

serves as a distinct feature for predicting student performance in previous studies~\cite{minn2018improving,minn2019dynamic,minn2022interpretable}. The difficulty of a problem $p_j$ is determined on a scale of 1 to 10. Problem difficulty $\textbf{Pd}(P_j)$ is calculated as:

\begin{equation} \label{equ:incorrratio} 
\textbf{Pd}(p_j) = 
\begin{cases}
\delta(p_{j}) & \text{if}\ \rvert N_j \lvert \geq 4\\
5 & \text{otherwise}
\end{cases}
\end{equation}

\begin{equation} \label{equ:catdelta}
\delta(p_{j}) = \left\lfloor  { \frac{{ \sum_i^{\lvert N_j \rvert }}{O_i(p_{j})}}{\lvert N_j \rvert}} \cdot 10   \right\rfloor
\end{equation}
and where $p_j$ represents the $j^{th}$ problem; $N_j$ is the set of students who attempted problem~$p_j$; $O_i({p_{j}})$ the outcome of the first attempt from student~$i$ to problem~$p_j$, 1~if successful, 0~otherwise. $\delta(p_{j})$ is a function that maps the \emph{average success rate} of problem~$p_j$ onto $10$ levels. 

\subsubsection{ Ability Profile: Learners trajectory}

Approaches like DKT-DSC and IKT~\cite{minn2018deep,minn2022interpretable} introduced the use of cluster IDs as an ability profile, first applied in deep learning-based models and later extended to Tree-Augmented Naive Bayes (TAN). While clustering offers a cost-efficient solution for deep learning-based models, it may sacrifice some predictive accuracy.

To address these challenges, we reformulate the \textit{ability profile} by directly introducing two types based on problem difficulty and skill correctness. This enhancement improves the model’s ability to track and adapt to student learning over time.

\begin{itemize}
    \item \textbf{Skill Correctness-based Ability Profile} is calculated to represent a student's cumulative ability in mastering the skill associated with the current problem. This measure provides valuable insights into a student's skill-specific strengths and weaknesses, enabling personalized learning interventions.

    \begin{equation}
    sr\_profile=R(x_{j})_{1:t}=\sum_{t=1}^{z}\frac{(x_{jt})}{|N_{jt}|},\label{equ:segcorr}
    \end{equation}%
    
    where $x_{jt}$ is the total correct outcome of attempt of skill~$x_{j}$ of the problem being correctly answered at timestamp $t$; $|N_{jt}|$ is the total number of practice attempts of skill~$x_{j}$ up to timestamp $t$;

    \item \textbf{Item Difficulty-based Ability Profile} is calculated to represent a student's cumulative ability in handling problems of varying difficulty levels. This metric helps in understanding how well a student performs across different difficulty tiers and provides a basis for adaptive instructional strategies tailored to their proficiency.

    \begin{equation}
    df\_profile=R(\textbf{Pd}(p_j))_{1:t}=\sum_{t=1}^{t}\frac{(\textbf{Pd}(p_{jt}))}{|N_{jt}|},\label{equ:segcorr}
    \end{equation}%
    
    where $\textbf{Pd}(p_{jt})$ is the total correct outcome of the attempts to problem difficulty level like current problem~$(p_j)$ being correctly answered until timestamp $t$; $|N_{jt}|$ is the total number of practice attempts of problem difficulty level like current problem~$\textbf{Pd}(p_j)$ up to timestamp $t$.
\end{itemize}

Instead of utilizing a cluster ID as an ability profile, as done in DKT-DSC and IKT, we employed a more refined ability profile derived from two key dimensions: item difficulty and skill correctness. This approach captures the cumulative ability of a student by leveraging their performance on problems categorized by difficulty levels and specific skills.

\section{Predicting student performance}

\begin{wrapfigure}{l}{0.40\textwidth}
	\centering 
        \includegraphics[width=0.40\textwidth]{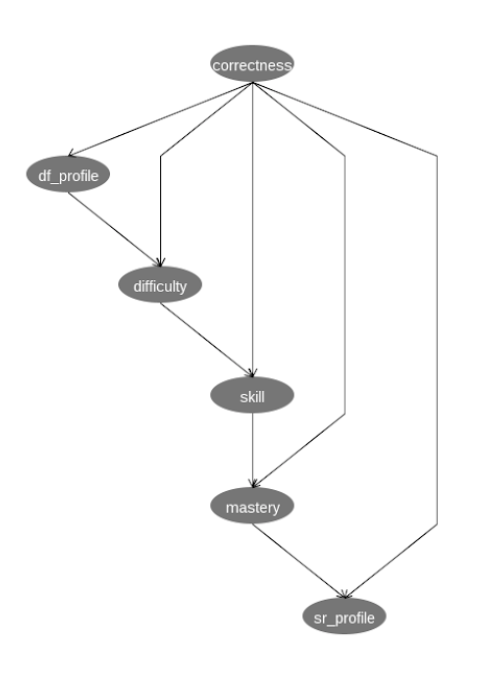}  
        \caption{Tree Structure of TAN for EIKT.}
	\label{fig:TAN} 
\end{wrapfigure}

Inspired by Interpretable Knowledge Tracing (IKT), we employ the Tree-Augmented Naive Bayes (TAN) method~\cite{friedman1997bayesian,minn2014efficient}. TAN extends the Naive Bayes network by designating the class node representing problem correctness as the root, causally connected to evidence nodes such as skill ID, skill mastery, ability profile, and problem difficulty. As shown in Figure \ref{fig:TAN},this approach maintains the directed acyclic graph (DAG) structure~\cite{fu2014towards,minn2016algorithm,sein2016accelerating}, ensuring efficient reasoning while retaining interpretability.

EIKT\footnote{ The source code and data for replicating our experiments are available at~\url{https://github.com/Simon-tan/IKT.git}} performs inference by using five meaningful  features $f_t$: skill ID (skill), skill mastery (mastery), skill correctness-based ability profile (sr\_profile), item Difficulty-based ability profile (df\_profile), and problem difficulty (difficulty) as evidenced at the current timestamp $t$ (for brevity we omit indexing all terms by student $i$, $s_t$, $P_j$ in Figure \ref{fig:TAN} and Equation \ref{equ:EIKT1}). 

\begin{equation}\label{equ:EIKT1}
P(correctness_t=y|f_t) = \frac{P(y) P(f_t|y)}{\sum_{y'}P(y') P(f_t|y')}
\end{equation}

\begin{align*}\label{equ:IKT2}
\textnormal{where } P(f_t|y)  = & P(\text{df\_profile}|y)  
P(\text{diffiuclty}|y,\text{df\_profile}) 
P(\text{skill}|y,\text{diffiuclty}) \\ &
P(\text{mastery}|y,\text{skill}) 
P(\text{sr\_profile}|y,\text{mastery}) 
\end{align*}

The class node (correctness) represents the predicted probability that the student would correctly answer the problem with the associated skill as described in Figure~\ref{fig:TAN}. We can achieve interpretation via the conditional probability tables of each node with their causal links~\cite{minn2023fast}.

The primary computational cost arises from extracting skill mastery using Bayesian Knowledge Tracing (BKT), while training the TAN model takes only a few seconds. Consequently, the overall training cost of our approach remains effectively equivalent to that of BKT, with the TAN modeling cost being negligible. Unlike IKT, our approach eliminates profile clustering, thereby reducing the computational expenses associated with clustering. This allows EIKT to significantly lower model complexity, training time, and computational costs compared to deep learning models while providing causal explanations for greater interpretability.

\section{Experiments}\label{sec:experimental-study}

We compare the next problem student performance prediction of our model with well-known KT models mentioned above:  BIRT~\cite{wilson2016back}, BKT~\cite{corbett1994knowledge}, PFA~\cite{pavlik2009performance}, DKT~\cite{piech2015deep}, DKT-DSC~\cite{minn2018deep}, DKVMN~\cite{zhang2017dynamic}, AKT-R~\cite{ghosh2020context} and IKT~\cite{minn2022interpretable}. To validate the proposed model, we conducted experiments using three public datasets from two distinct computer-based tutoring scenarios in educational settings: {ASSISTments:} 1)ASSISTments 2009-2010 (Skill Builder) 2)ASSISTments 2012-2013~\cite{feng2009addressing} \footnote{\url{https://sites.google.com/site/assistmentsdata/}} and Cognitive Tutor Algebra 2005-2006 (KDD Cup 2010) dataset on PSLC DataShop\footnote{\url{https://pslcdatashop.web.cmu.edu/KDDCup/downloads.jsp}}. For all datasets, only the first correct attempts to original problems are considered in our experiment. We remove data with missing values for skills and problems with duplicate records. We used five-fold cross-validation to evaluate all datasets, splitting each fold into 80\% training and 20\% testing students. To the best of our knowledge, these are among the most well known publicly available knowledge tracing datasets.

\subsubsection{Results}
\begin{table}
    \centering
    \renewcommand{\arraystretch}{1.3} 
    \begin{tabularx}{\textwidth}{c|c||*{9}{X}} 
        \hline
        Type & Datasets & BIRT & PFA & BKT & DKT & DKT-DSC & DKV-MN & AKT-R & IKT & \textbf{EIKT} \\
        \hline
        \multirow{4}{*}{AUC} 
        & ASS-09 & 0.750 & 0.701 & 0.651 & 0.721 & 0.735 & 0.710 & 0.767 & \underline{0.797} & \textbf{0.801} \\
        & ASS-12 & 0.744 & 0.672 & 0.623 & 0.713 & 0.721 & 0.707 & \textbf{0.777} & 0.767 & \underline{0.773} \\
        & Algebra & 0.812 & 0.754 & 0.642 & 0.784 & 0.792 & 0.780 & \underline{0.845} & \underline{0.845} & \textbf{0.848} \\
         \cmidrule(lr){2-11}
        & Average & 0.768 & 0.709 & 0.638 & 0.739 & 0.749 & 0.732 & 0.796 & \underline{0.802} & \textbf{0.806} \\
        \hline

        \multirow{4}{*}{RMSE} 
        & ASS-09 & 0.440 & 0.454 & 0.471 & 0.450 & 0.434 & 0.451 & {0.423} & \underline{0.412} & \textbf{0.411} \\
        & ASS-12 & 0.441 & 0.440 & 0.510 & 0.430 & 0.427 & 0.430 & \textbf{0.409} & \underline{0.412} & \underline{0.412} \\
        & Algebra & 0.374 & 0.391 & 0.440 & 0.380 & 0.373 & 0.380 & \underline{0.357} & \underline{0.358} & \textbf{0.356} \\
          \cmidrule(lr){2-11}
        & Average & 0.418 & 0.428 & 0.473 & 0.420 & 0.411 & 0.420 & 0.396 & \underline{0.395} & \textbf{0.394} \\
         \hline
    \end{tabularx}
    
        \small 
        Best scores are in bold, second-best scores are underlined.
    
    \caption{Average AUC and RMSE result for all tested models across datasets.}
    \label{tab:exp1}
\end{table}
The EIKT structure results in a student model with better explanation with causal relations and higher predictive performance. The results in Tables \ref{tab:exp1} demonstrate that EIKT outperforms significantly the well-known KT models in all tested datasets. When we compare EIKT with our second-best performer IKT and AKT-R (scores are underlined in table \ref{tab:exp1}). EIKT shows slightly better performance than IKT but has superior performance compared to other models tested in our experiments.  the improvement is ranging from 0.12\% to 0.78\% in terms of AUC. When we compare in terms of RMSE, it still keeps almost the same performance as IKT but shows better performance than other models. So EIKT shows better performance than any other method in both AUC and RMSE (except AKT-R on the ASS-12 dataset).

\section{Conclusion}\label{sec:conclusion}

Unlike the cluster-ID used as an ability profile in IKT, we introduce two enhanced ability profiles that improve student performance prediction while preserving model interpretability. By incorporating these features into the Tree-Augmented Naïve Bayes (TAN) structure with explicit causal relationships, we develop a Knowledge Tracing (KT) model that achieves superior predictive accuracy without relying on extensive parameters or complex architectures.

We propose a causal probabilistic student model, EIKT, which integrates five key features: skill ID, student skill mastery (probability of mastering a skill), two ability profiles (based on problem difficulty and skill correctness), and problem difficulty. Unlike deep learning-based KT models that rely on hidden states, numerous parameters, and complex architectures, EIKT leverages explicit, interpretable features derived from skills, problems, and students, enhancing predictive accuracy and causal interpretability. By eliminating profile clustering, EIKT further reduces the computational expenses required in IKT, making it even more efficient. Experiments on three public datasets demonstrate that EIKT outperforms established KT models, including deep learning approaches, while maintaining significantly lower computational costs. The integration of carefully engineered features captures greater data variance, while the TAN framework provides causal insights. By incorporating skill mastery, ability profiles, and problem difficulty, EIKT effectively accounts for student variability, enabling more personalized and accurate predictions.

While our primary focus is on prediction accuracy, we also highlight the importance of understanding the causal effects of explainable features to inform adaptive instruction in personalized learning systems. Future work will explore knowledge acquisition and learning behaviors to further optimize human learning in evolving educational environments.
\bibliographystyle{splncs04}
\bibliography{bibliography}

\begin{thebibliography}{10}
\providecommand{\url}[1]{\texttt{#1}}
\providecommand{\urlprefix}{URL }
\providecommand{\doi}[1]{https://doi.org/#1}

\bibitem{corbett1994knowledge}
Corbett, A.T., Anderson, J.R.: Knowledge tracing: Modeling the acquisition of procedural knowledge. User modeling and user-adapted interaction  \textbf{4}(4),  253--278 (1994)

\bibitem{feng2009addressing}
Feng, M., Heffernan, N., Koedinger, K.: Addressing the assessment challenge with an online system that tutors as it assesses. User Modeling and User-Adapted Interaction  \textbf{19}(3),  243--266 (2009)

\bibitem{friedman1997bayesian}
Friedman, N., Geiger, D., Goldszmidt, M.: Bayesian network classifiers. Machine learning  \textbf{29}(2),  131--163 (1997)

\bibitem{fu2014towards}
Fu, S., Minn, S., Desmarais, M.C.: Towards the efficient recovery of general multi-dimensional bayesian network classifier. In: Machine Learning and Data Mining in Pattern Recognition: 10th International Conference, MLDM 2014, St. Petersburg, Russia, July 21-24, 2014. Proceedings 10. pp. 16--30. Springer (2014)

\bibitem{fu2013recommender}
Fu, S., Zhang, Y., et~al.: On the recommender system for university library. International Association for Development of the Information Society  (2013)

\bibitem{ghosh2020context}
Ghosh, A., Heffernan, N., Lan, A.S.: Context-aware attentive knowledge tracing. In: Proceedings of the 26th ACM SIGKDD International Conference on Knowledge Discovery \& Data Mining. pp. 2330--2339 (2020)

\bibitem{minn2020bkt}
Minn, S.: Bkt-lstm: Efficient student modeling for knowledge tracing and student performance prediction. arXiv preprint arXiv:2012.12218  (2020)

\bibitem{minn2022ai}
Minn, S.: Ai-assisted knowledge assessment techniques for adaptive learning environments. Computers and Education: Artificial Intelligence  \textbf{3},  100050 (2022)

\bibitem{minn2025understanding}
Minn, S.: Understanding the landscape of ai-assisted adaptive learning systems in education. arXiv preprint  (2025)

\bibitem{minn2016refinement}
Minn, S., Desmarais, M.C., Fu, S.: Refinement of a q-matrix with an ensemble technique based on multi-label classification algorithms. In: European Conference on Technology Enhanced Learning. pp. 165--178. Springer (2016)

\bibitem{minn2019dynamic}
Minn, S., Desmarais, M.C., Zhu, F., Xiao, J., Wang, J.: Dynamic student classiffication on memory networks for knowledge tracing. In: Pacific-Asia Conference on Knowledge Discovery and Data Mining. pp. 163--174. Springer (2019)

\bibitem{minn2016algorithm}
Minn, S., Fu, S., Lv, T.: Algorithm for exact recovery of bayesian network for classification. Application Research of Computers  \textbf{33}(05),  1327--1334 (2016)

\bibitem{sein2016accelerating}
Minn, S., Fu, S.k.: Accelerating structure learning of bayesian network. Computer Science  (2016)

\bibitem{minn2014efficient}
Minn, S., Fu, S., Desmarais, M.C.: Efficient learning of general bayesian network classifier by local and adaptive search. In: 2014 International Conference on Data Science and Advanced Analytics (DSAA). pp. 385--391. IEEE (2014)

\bibitem{minn2023fast}
Minn, S., Shunkai, F.: Fast \& efficient learning of bayesian networks from data: Knowledge discovery and causality. In: 2023 IEEE International Conference on Data Mining Workshops (ICDMW). pp. 966--975. IEEE (2023)

\bibitem{minn2022interpretable}
Minn, S., Vie, J.J., Takeuchi, K., Kashima, H., Zhu, F.: Interpretable knowledge tracing: Simple and efficient student modeling with causal relations. In: Proceedings of the AAAI conference on artificial intelligence. vol.~36, pp. 12810--12818 (2022)

\bibitem{minn2018deep}
Minn, S., Yu, Y., Desmarais, M.C., Zhu, F., Vie, J.J.: Deep knowledge tracing and dynamic student classification for knowledge tracing. In: 2018 IEEE International conference on data mining (ICDM). pp. 1182--1187. IEEE (2018)

\bibitem{minn2018improving}
Minn, S., Zhu, F., Desmarais, M.C.: Improving knowledge tracing model by integrating problem difficulty. In: 2018 IEEE International Conference on Data Mining Workshops (ICDMW). pp. 1505--1506. IEEE (2018)

\bibitem{pavlik2009performance}
Pavlik, P.I., Cen, H., Koedinger, K.R.: Performance factors analysis--a new alternative to knowledge tracing. In: 14th International Conference on Artificial Intelligence in Education (2009)

\bibitem{piech2015deep}
Piech, C., Bassen, J., Huang, J., Ganguli, S., Sahami, M., Guibas, L.J., Sohl-Dickstein, J.: Deep knowledge tracing. In: Advances in neural information processing systems. pp. 505--513 (2015)

\bibitem{wilson2016back}
Wilson, K.H., Karklin, Y., Han, B., Ekanadham, C.: Back to the basics: Bayesian extensions of irt outperform neural networks for proficiency estimation. arXiv preprint arXiv:1604.02336  (2016)

\bibitem{zhang2017dynamic}
Zhang, J., Shi, X., King, I., Yeung, D.Y.: Dynamic key-value memory networks for knowledge tracing. In: Proceedings of the 26th international conference on World Wide Web. pp. 765--774 (2017)

\end{thebibliography}
\end{document}